\documentclass[prl,twocolumn]{revtex4}
\usepackage{dcolumn}
\usepackage{multirow}
\usepackage{graphicx}
\usepackage{amssymb}
\usepackage{bm}
\usepackage{hyperref}
\usepackage{epstopdf}
\usepackage{color}
\usepackage{mathrsfs}
\usepackage{amsmath,amssymb,amsthm}
\usepackage{rotating}
\usepackage{sverb, longtable}
\usepackage{subfigure}

\usepackage[graphicx]{realboxes}
\usepackage{adjustbox}
\begin{document}

\title{Evidence for Dynamical Dark Matter}
\author{Deng Wang}
\email{dengwang@ific.uv.es}
\affiliation{Instituto de F\'{i}sica Corpuscular (CSIC-Universitat de Val\`{e}ncia), E-46980 Paterna, Spain}

\begin{abstract}
The nature of dark matter is one of the most fundamental questions in cosmology. Using the cosmic microwave background (CMB), type Ia supernova (SN) and DESI's new measurements of baryon acoustic oscillations (BAO), we find the robust $\sim2\,\sigma$ evidences of the evolution of dark matter in the dynamical dark matter (DDM) model, $\omega_{dm}(a)=\omega_{dm0}+\omega_{dma}(1-a)$. 
Based on CMB data, we find a very strong linear relation $\omega_{dma}=-\omega_{dm0}$, inducing the single-parameter DDM model, $\omega_{dm}(a)=\omega_{dm}a$, where the $\sim2\,\sigma$ DDM evidences is well captured and even strengthened. We demonstrate that there are beyond $2\,\sigma$ evidences of the coexistence of DDM and dynamical dark energy using the combinations of CMB, DESI BAO and Pantheon+ SN data.
In such models, at a beyond $5\,\sigma$ confidence level, we verify that the universe remains in a matter-dominated state for a substantial period in the past, accelerate in the distant future and finally becomes completely dominated by dark matter. We propose that the ultimate fate of the universe is the ``Super Rip'' induced by dark matter with an extremely negative pressure. 
Our findings fundamentally challenge the prevailing understanding of cosmic acceleration and deepen our insight into the universe's evolution.

\end{abstract}
\maketitle



{\it Introduction.} Dark matter (DM), an invisible and pervasive constituent of the universe, has captivated scientists for nearly a century. Its existence is inferred through its gravitational influence on the visible matter, radiation, and the large-scale structure of the universe. Despite its elusive nature, DM is believed to constitute approximately 85\% of the total matter in the universe, playing a crucial role in shaping its evolution and dynamics \cite{Bertone:2004pz,Bertone:2016nfn,Weinberg:2013agg}.

The concept of DM emerged from discrepancies between the observed dynamics of astrophysical systems and the predictions of Newtonian gravity based on the visible matter alone. Early hints of DM came from the investigations of galaxy clusters and galactic rotation curves, which revealed that the visible matter could not account for the observed gravitational effects \cite{Zwicky:1933gu,Rubin:1970zza}. The theoretical framework for DM has evolved significantly over the decades. The Cold Dark Matter (CDM) model, proposed in the 1980s, suggests that DM particles are ``cold,'' indicating that they move slowly compared to the speed of light. This scenario successfully explains the large-scale structure of the universe, including the formation and distribution of galaxies and galaxy clusters \cite{Blumenthal:1984bp}. Alternative models, such as Warm Dark Matter (WDM) and Hot Dark Matter (HDM), have also been proposed. WDM particles are hypothesized to move at relativistic speeds but are still massive enough to contribute to structure formation. HDM particles, on the other hand, move at speeds close to the speed of light and are less effective at forming structures \cite{Dodelson:1993je}.

Observational evidences for DM comes from a variety of sources, including gravitational lensing, the CMB radiation \cite{COBE:1992syq}, and the dynamics of galaxy clusters and individual galaxies\cite{Weinberg:2013agg}. Gravitational lensing, the bending of light by massive objects, provides a direct probe of the distribution of DM in the universe. Precision CMB measurements of the WMAP \cite{WMAP,WMAP:2012fli} and Planck \cite{Planck,Planck:2018vyg} satellites have revealed that DM constitutes approximately 27\% of the total mass-energy content of the universe.

The search for DM has led to numerous detection efforts, including direct detection experiments, indirect detection methods and collider searches. Direct detection experiments aim at observing the scattering of DM particles off atomic nuclei in the laboratory. Detectors such as XENON \cite{XENON,XENON100:2012itz}, LUX \cite{LUX,LUX:2016ggv}, and PandaX \cite{PANDAX,PandaX:2014mem} employ large volumes of liquid xenon or other materials to search for the tiny signals produced by DM interactions \cite{XENON:2018voc}. Indirect detection methods look for the products of DM decay or annihilation, such as gamma rays, neutrinos, and antimatter. Space-based telescopes like Fermi-LAT\cite{FERMI-LAT} and ground-based observatories like HESS \cite{HESS} and MAGIC \cite{MAGIC,MAGIC:2016xys,Profumo:2017obk} search for these signals from astrophysical sources \cite{Klasen:2015uma,Gaskins:2016cha}. Collider searches, such as those conducted at the Large Hadron Collider (LHC) \cite{LHC}, aim at producing DM particles and observe their signatures in the form of missing energy and momentum. These experiments provide a complementary approach to direct and indirect detection methods, offering the potential to discover new physics beyond the Standard Model \cite{Bertone:2004pz,Bertone:2016nfn}. Despite decades of research, DM remains undetected, posing significant challenges for both theoretical and experimental physics. The weak interactions of DM with ordinary matter, the vast parameter space of possible DM candidates, and the need for highly sensitive detectors are among the primary obstacles in the quest to unravel the mysteries of DM \cite{Bertone:2018krk}. Future efforts in DM research will focus on improving detector technologies, exploring new theoretical models, and combining data from multiple sources to enhance sensitivity. The upcoming generation of experiments, including the Lux-Zeplin (LZ) \cite{LZ:2019sgr} direct detection experiment and the Cherenkov Telescope Array (CTA) \cite{CTAConsortium:2017dvg} for indirect detection, promise to push the boundaries of our understanding of DM.

During the past two decades, accompanying with the discovery of dark energy (DE) from two SN teams \cite{SupernovaSearchTeam:1998fmf,SupernovaCosmologyProject:1998vns}, the CMB \cite{Planck:2018vyg,ACT:2025fju,SPT-3G:2022hvq,WMAP:2003elm,Planck:2013pxb}, baryon acoustic oscillations (BAO) \cite{SDSS:2005xqv,2dFGRS:2005yhx,Beutler:2011hx,eBOSS:2020yzd,deCarvalho:2017xye,eBOSS:2017cqx,DESI:2024mwx,DESI:2024uvr,DESI:2024lzq}, weak gravitational lensing \cite{Heymans:2012gg,Hildebrandt:2016iqg,Planck:2018lbu,DES:2017qwj}, galaxy clustering \cite{DES:2017myr,DES:2021wwk}, cluster abundance \cite{DES:2025xii} and SN observations \cite{SDSS:2014iwm,SupernovaCosmologyProject:2011ycw,Pan-STARRS1:2017jku,Brout:2022vxf,Rubin:2023ovl,DES:2024jxu} have further deepened our understanding of cosmic acceleration and confirmed the validity of $\Lambda$CDM. However, it confronts at least two challenges, i.e., the cosmological constant conundrum \cite{Weinberg:1988cp,Carroll:2000fy,Peebles:2002gy,Padmanabhan:2002ji} and the coincidence problem \cite{Steinhardt:1997}, while suffering from the emergent cosmic tensions of the Hubble constant ($H_0$) and the matter fluctuation amplitude ($S_8$) \cite{DiValentino:2020vhf,DiValentino:2020zio,DiValentino:2020vvd,Abdalla:2022yfr,DiValentino:2025sru}. Logically, these progresses compel theorists to explore new physics beyond the $\Lambda$CDM model in order to address these issues (see \cite{Abdalla:2022yfr,DiValentino:2025sru} for details).

Recently, the DESI collaboration has reported substantial evidences for the Dynamical Dark Energy (DDE) based on the BAO measurements from their second data release (DR2) \cite{DESI:2025zgx,DESI:2025zpo}. This implies that the nature of DE is likely evolving over time \cite{DESI:2025fii,Wang:2025bkk}. However, this is obtained based on the traditional assumption that DM is absolutely cold, i.e., the equation of state (EoS) of DM is zero. Therefore, it is natural to raise a similar question: whether DM also evolve over time? Since DM and DE can simultaneously affect the background dynamics and structure formation of the universe, they are closely related to each other.
Therefore, an even more interesting question is whether both DM and DE are dynamical in the dark sector of the universe?

Note that, based on the theoretical motivation proposed in \cite{Hu:1998kj}, although the scenario of DM with non-constant EoS has been studied in \cite{Kumar:2012gr,Kumar:2019gfl}, the incomplete parameter space of DM EoS limits the exploration of possible new physics. Besides using new data, we extend the theoretical parameter space of DM EoS to the full space. Employing individual datasets CMB, DESI and Pantheon+, we find $\sim2\,\sigma$ DDM evidences. Using the data combinations of CMB and low-$z$ probes, we find beyond $2\,\sigma$ signals of the coexistence of DDM and DDE.

{\it Model.} In the framework of general relativity \cite{Einstein:1916vd}, considering a homogeneous and isotropic universe, the Friedmann equations read as $H^2=(8\pi G\rho)/3$ and $\ddot{a}/a=-4\pi G(\rho+3p)/3$, where $H$ is the cosmic expansion rate at a scale factor $a$ and $\rho$ and $p$ are the mean energy densities and pressures of different species including baryons, DM and DE in the late universe. Combining two Friedmann equations \cite{Friedman:1922kd} and assuming the DM EoS $\omega_{dm}(a)=\omega_{dm0}+\omega_{dma}(1-a)$ and the Chevallier-Polarski-Linder (CPL) DE EoS $\omega(a)=\omega_0+\omega_a(1-a)$ \cite{Chevallier:2000qy,Linder:2002et}, the normalized Hubble parameter $E(a)\equiv H(a)/H_0$ is written as
\begin{equation}
E(a)=\left[\Omega_{b}a^{-3}+\Omega_{\rm DM}(a)+\Omega_{\rm DE}(a)\right]^{\frac{1}{2}}, \label{eq:ez}
\end{equation}
where $\Omega_{\rm DM}(a)=\Omega_{dm}a^{-3(1+\omega_{dm0}+\omega_{dma})}\mathrm{e}^{3\omega_{dma}(a-1)}$ and
$\Omega_{\rm DE}(a)=\Omega_{\rm de}a^{-3(1+\omega_0+\omega_a)}\mathrm{e}^{3\omega_a(a-1)}$, where $\Omega_b$, $\Omega_{\rm dm}$ and $\Omega_{\rm de}$ ($=1-\Omega_{dm}-\Omega_{b}$) are today's baryon, DM and DE fractions, respectively. It reduces to $\Lambda$CDM when $\omega_{dm0}=\omega_{dma}=\omega_a=0$ and $\omega_0=-1$.

\begin{figure}[h]
	\centering
	\includegraphics[scale=0.5]{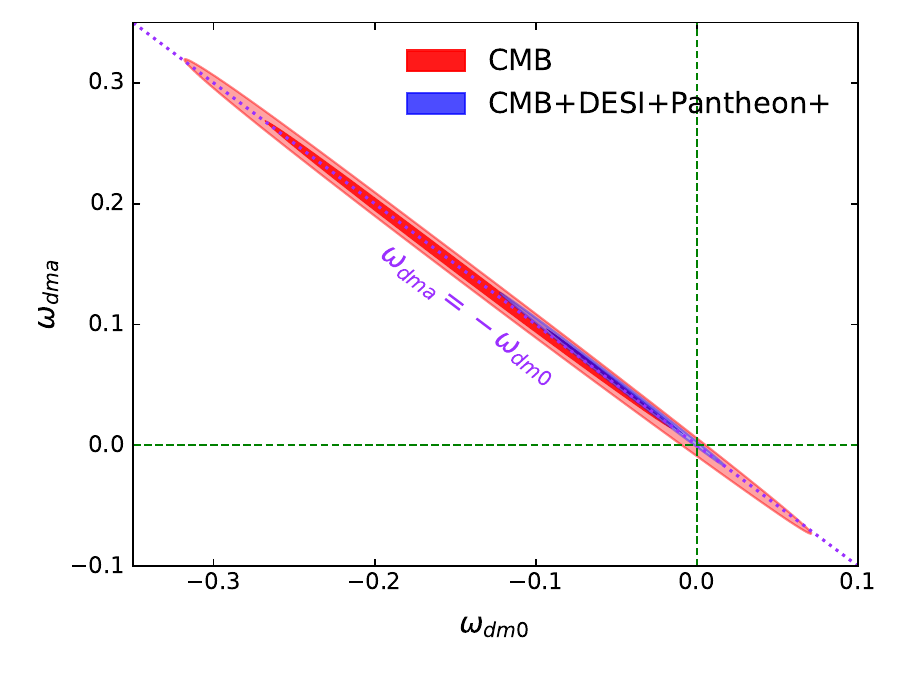}
	\caption{Two-dimensional posterior distributions of the parameter pair ($\omega_{dm0}$, $\omega_{dma}$) from CMB and CMB+DESI+Pantheon+ datasets in the DDM model. The cross point of green dashed lines denotes $\Lambda$CDM, while the purple dotted line represents the linear relation $\omega_{dma}=-\omega_{dm0}$. }\label{f1}
\end{figure}

{\it Data and methods.} We take the Planck 2018 high-$\ell$ \texttt{plik} temperature (TT) likelihood at multipoles  $30\leqslant\ell\leqslant2508$, polarization (EE) and their cross-correlation (TE) data at $30\leqslant\ell\leqslant1996$, and the low-$\ell$ TT \texttt{Commander} and \texttt{SimAll} EE likelihoods at $2\leqslant\ell\leqslant29$ \cite{Planck:2019nip}. We use conservatively the Planck lensing likelihood from \texttt{SMICA} maps at $8\leqslant\ell \leqslant400$ \cite{Planck:2018lbu}. We use 13 BAO measurements from DESI DR2 including the BGS, LRG1, LRG2, LRG3+ELG1, ELG2, QSO and Ly$\alpha$ samples at the effective redshifts $z_{\rm eff}=0.295$, 0.51, 0.706, 0.934, 1.321, 1.484 and $2.33$, respectively \cite{DESI:2025zgx,DESI:2025fii,DESI:2025zpo}. We adopt the Pantheon+ SN sample consisting of 1701 data points from 18 different surveys in $z\in[0.00122, 2.26137]$ \cite{Brout:2022vxf}. Hereafter, we denote CMB, DESI and SN as ``C'', ``D'' and ``S'', respectively.   

We take the Boltzmann solver \texttt{CAMB} \cite{Lewis:1999bs} to compute the background evolution and theoretical power spectra. To perform the Bayesian analysis, we use the Monte Carlo Markov Chain (MCMC) method to infer the posterior distributions of model parameters using \texttt{CosmoMC} \cite{Lewis:2002ah,Lewis:2013hha}. We assess the convergence of MCMC chains via the Gelman-Rubin criterion $R-1\lesssim 0.01$ \cite{Gelman:1992zz} and analyze them using \texttt{Getdist} \cite{Lewis:2019xzd}.

We use the following uniform priors for model parameters: the baryon fraction $\Omega_bh^2 \in [0.005, 0.1]$, cold DM fraction $\Omega_ch^2 \in [0.001, 0.99]$, acoustic scale at the recombination epoch $100\theta_{\rm MC} \in [0.5, 10]$, spectral index $n_s \in [0.8, 1.2]$, amplitude of the primordial power spectrum $\ln(10^{10}A_s) \in [2, 4]$, optical depth $\tau \in [0.01, 0.8]$, present-day DM EoS $\omega_{dm0} \in [-2, 2]$, amplitude of DM evolution $\omega_{dma} \in [-2, 2]$ and $\omega_{dm} \in [-2, 2]$, today's DE EoS $\omega_0 \in [-15, 20]$ and amplitude of DE evolution $\omega_a \in [-30, 10]$. For DESI, Pantheon+ and their combination, we use $\Omega_b \in [0.001, 0.1]$, $\Omega_c \in [0.001, 0.99]$, $\omega_{dm0} \in [-10, 10]$ and $\omega_{dma} \in [-10, 10]$.
Below, we refer the single-parameter DDM and DDM plus DDE scenarios as ``SDDM'' and ``DDME'', respectively. 

\begin{figure}
	\centering
	\includegraphics[scale=0.55]{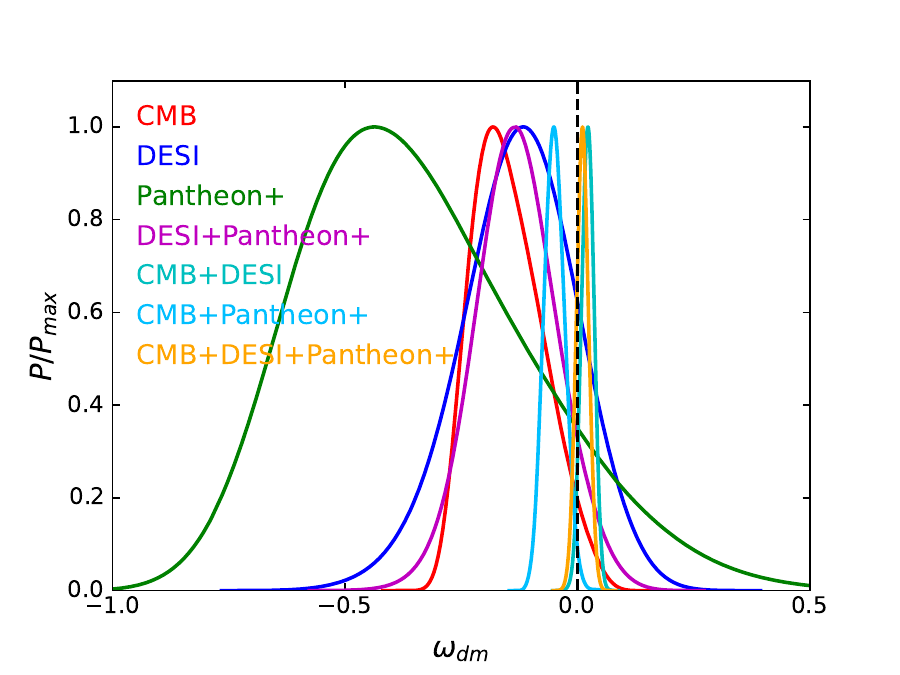}
	\caption{One-dimensional posterior distributions of the parameter $\omega_{dm}$ from various datasets in the SDDM model. The black dashed line denotes $\Lambda$CDM.}\label{f2}
\end{figure}

\begin{figure}
	\centering
	\includegraphics[scale=0.42]{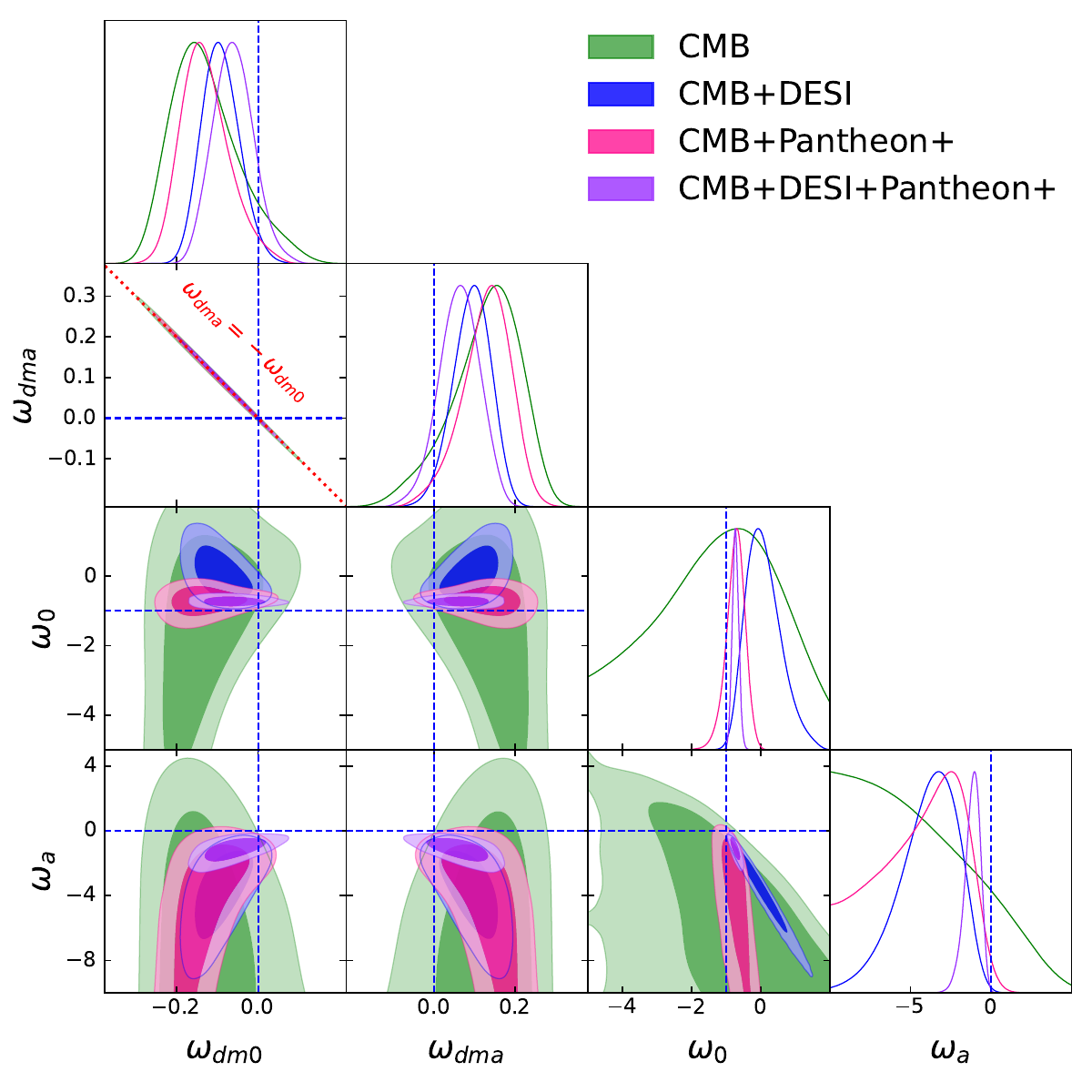}
	\caption{One-dimensional and two-dimensional posterior distributions of the DDM and DDE EoS parameters from various datasets in the DDME model. The blue dashed lines are $\omega_{dm0}=\omega_{dma}=\omega_a=0$ and $\omega_0=-1$, while the red dotted line denotes the linear relation $\omega_{dma}=-\omega_{dm0}$.}\label{f3}
\end{figure}

{\it DDM evidence.} In the DDM model, CMB and CDP gives $\omega_{dm0}=-0.139^{+0.076}_{-0.091}, \, -0.051\pm 0.029$ and $\omega_{dma}=0.138^{+0.092}_{-0.077}, \, 0.053\pm 0.029$, indicating a $1.96\,\sigma$ and $1.85\,\sigma$ evidence of evolving DM, respectively. This means that both the early-time probe CMB alone and its combination with two late-time probes DESI and Pantheon+ prefer the DDM at approximately $2\,\sigma$ level (see Fig.\ref{f1}). DESI gives $\omega_{dma}=0.56^{+0.47}_{-0.29}$ a $1.35\,\sigma$ DDM hint, but Pantheon+ cannot provide a clue of DDM due to its weak constraining power (see Tab.\ref{t1}). In pairwise data combinations, only CP provides a $1.05\,\sigma$ hint of DDM, while CD just gives a slight preference of $\omega_{dma}>0$. The late-time combination DP shows no deviation from $\Lambda$CDM. Interestingly, based on CMB data, we find there is a very strong linear relation $\omega_{dma}=-\omega_{dm0}$ between two DM EoS parameters. Therefore, we propose the SDDM model
\begin{equation}
\omega_{dm}(a)=\omega_{dm}a, \label{eq:eos}
\end{equation}
where $\omega_{dm}$ is the sole parameter depicting the redshift evolution of DM EoS. We find that CMB, DESI and Pantheon+ give $\omega_{dm}=-0.149^{+0.061}_{-0.092}, \, -0.12^{+0.13}_{-0.12}$ and $-0.35^{+0.20}_{-0.29}$, implying a $2.33\,\sigma$, $\sim1\sigma$ and $1.44\,\sigma$ DDM evidence (see Fig.\ref{f2}), respectively. The joint constraint from two late-time probes DP gives $1.49\,\sigma$ hint of $\omega_{dm}<0$, while CP provides a $2.26\,\sigma$ DDM evidence (see Tab.\ref{t1}). However, CD and CDP give $1.57\,\sigma$ and $0.85\,\sigma$ evidences of $\omega_{dm}>0$. The origin of these anomalous positive values of $\omega_{dm}$ is that DESI plus the high-precision information of comoving sound horizon $r_d$ from CMB leads to a higher $H_0$ than CMB. Consequently, CD and CDP provide problematic constraints on $\omega_{dm}$, while constraints from C, D, P, DP and CP are more robust. It is easy to see that $\omega_{dm}<0$ in the SDDM scenario perfectly captures the information from CPL DDM parameter space in the second quadrant and even gives stronger signals of non-constant DM EoS.

\begin{table*}[t!]
	\renewcommand\arraystretch{1.5}
	\caption{Mean values and $1\,\sigma$ (68\%) errors of free parameters from various datasets in the DDM, SDDM and DDME models. We quote $2\,\sigma$ (95\%) upper limits for parameters with weak constraints, while using the symbols ``$\bigstar$'' to denote unconstrained parameters by data. }
	\setlength{\tabcolsep}{1.3mm}{
		\begin{tabular} { c |c| c |c| c|c |c|c |c}
			\hline
			\hline
			
			\multicolumn{2}{c|}{Data}                & C      & D          & P    &DP       & CD    & CP & CDP                  \\
			\hline
			\multirow{2}{1cm}{DDM}  & $\omega_{dm0}$  &$-0.139^{+0.076}_{-0.091}  $ & $-0.43^{+0.21}_{-0.33}     $ & $-0.04^{+0.40}_{-0.66}     $ & $-0.04^{+0.10}_{-0.15}     $ & $-0.020\pm 0.032           $ & $-0.046\pm 0.043           $ & $-0.051\pm 0.029           $                                                    \\
			\cline{2-9}
			&    $\omega_{dma}$  & $0.138^{+0.092}_{-0.077}   $ & $0.56^{+0.47}_{-0.29}      $ & $-0.29^{+1.40}_{-0.79}      $ & $0.00^{+0.24}_{-0.18}      $ & $0.021\pm 0.032            $ & $0.046\pm 0.044            $ & $0.053\pm 0.029$                                                  \\

			\hline
			\multirow{1}{1cm}{SDDM}  &     $\omega_{dm}$  & $-0.149^{+0.061}_{-0.092}  $ & $-0.12^{+0.13}_{-0.12}     $ & $-0.35^{+0.20}_{-0.29}     $ & $-0.131\pm 0.088           $ & $0.022\pm 0.014            $ & $-0.050^{+0.021}_{-0.024}  $ & $0.011\pm 0.013            $                                                 \\
			
			\hline
			\multirow{4}{1cm}{DDME}  &     $\omega_{dm0}$  & $-0.127^{+0.061}_{-0.100}   $ & $-0.72\pm 0.23$ & $-0.76^{+1.10}_{-0.73}      $ & $-0.99^{+1.6}_{-1.1}$ & $-0.091^{+0.042}_{-0.050}  $ & $-0.128^{+0.047}_{-0.068}  $ & $-0.061\pm 0.050           $                                         \\
			\cline{2-9}
			&     $\omega_{dma}$  & $0.127^{+0.100}_{-0.062}    $ & $0.93^{+0.15}_{-0.19} $ & $< 3.39      $ & $-0.2\pm 4.0 $ & $0.093^{+0.051}_{-0.043}   $ & $0.128^{+0.069}_{-0.049}   $ & $0.062\pm 0.051            $                                           \\
			\cline{2-9}                        
			&     $\omega_{0}$   & $\bigstar$                          & $< 1.23$                          & $-1.5^{+2.5}_{-1.3}$                          & $-0.68\pm 0.23  $ & $0.09^{+0.39}_{-0.60}      $ & $-0.73^{+0.30}_{-0.24}     $ & $-0.729\pm 0.094           $                                             \\	
			\cline{2-9}	                        		                        
			&     $\omega_{a}$    & $< 1.81                    $ & $\bigstar$     & $\bigstar$    & $0.75^{+0.60}_{-0.51}                     $ & $-3.9^{+2.2}_{-1.3}         $ & $-4.5^{+3.5}_{-2.0}        $ & $-1.15^{+0.56}_{-0.38}     $                                               \\
			
			\hline
			\hline
	\end{tabular}}
	\label{t1}
\end{table*}

{\it The coexistence of DDM and DDE.} Regarding the late universe, the contributions of DM and DE to the background evolution of the universe are comparable. Throughout the entire history of the universe, DM plays a more significant role than DE in affecting the cosmic evolution. Although the DESI's claim of DDE has gained much attention assuming CDM with zero EoS, we have to question whether DDM and DDE coexist in the dark sector of the universe. Confronting the DDME model with observations, we find that CMB gives $\omega_{dma}=0.127^{+0.100}_{-0.062}$ indicating a $1.82\,\sigma$ evidence of DDM. Interestingly, DESI gives $\omega_{dma}=0.93^{+0.15}_{-0.19}$ and $\omega_{dm0}=-0.72\pm 0.23$ suggesting a $2.12\,\sigma$ DDM evidence and a $3.1\,\sigma$ signal of the negative pressure of today's DM (see Tab.\ref{t1}). Pantheon+ cannot provide a good constraint due to its limited constraining power. Note that DP gives a weaker constraint than DESI alone, since Pantheon+ significantly affect the DM abundance preferred by DESI. Furthermore, CD, CP and CDP exhibit a $2.21\,\sigma$, $2.42\,\sigma$ and $1.39\,\sigma$ evidence of DDM, respectively. The addition of DESI to CP leads to a larger $H_0$, reducing $\Omega_m$, and consequently suppresses the significance of $\omega_{dma}>0$ via the positive correlation between $\Omega_m$ and $\omega_{dma}$.

Similar to the DDE-only case \cite{Wang:2025bkk}, CMB gives a $\sim2\,\sigma$ evidence of DDE in the DDME model (see Fig.\ref{f3}). CD, CP and CDP provide, respectively, a $2.23\,\sigma$, $1.8\,\sigma$ and $2.44\,\sigma$ DDE evidence and a $2.22\,\sigma$, $\sim1\,\sigma$ and $2.89\,\sigma$ detection of present-day quintessence-like DE \cite{Ratra:1987rm,Wetterich:1987fm}. Interestingly, CD, CP and CDP all support the coexistence of DDM and DDE at a beyond $2\,\sigma$ level in the planes of $\omega_{dm0}-\omega_0$, $\omega_{dm0}-\omega_a$, $\omega_{dma}-\omega_0$ and $\omega_{dma}-\omega_a$. This implies that natures of both DM and DE are likely to be dynamical. Moreover, the very strong linear relation $\omega_{dma}=-\omega_{dm0}$ is confirmed in the DDME scenario. This seemingly reveals that the nature of DM is not only dynamical but also linear on cosmic scales at the phenomenological level.


{\it Destiny+.} In the DDE-only model, the ``Big Stall'' \cite{Wang:2025owe} predicts that the universe ultimately comes to a halt and will not reach a state of maximum entropy. However, this is not the case in the DDM, SDDM and DDME scenarios. Since only CMB can constrain the baryon fraction and perturbation behavior of DM, it is the most reliable tool to probe the composition and evolution of the universe. We show our results in the supplementary material (SM). In DDM and SDDM, using CMB, we cannot determine whether today's universe is accelerating within $1\,\sigma$ level (see SM). At a beyond $5\,\sigma$ level, the universe remains in a matter-dominated state for a substantial period in the past, accelerate in the distant future and becomes completely dominated by DM, i.e., $\Omega_m=1$. The universe possibly experiences an extremely accelerated expansion when reaching its end. Ultimately, stars and galaxies will be abruptly disrupted at an extraordinarily high velocity. We call this fate as the ``Super Rip'', which is produced by DM with an extremely negative pressure. Note that this extremely negative DM EoS is inconsistent with the principle of causality, which requires that information or energy cannot travel faster than the speed of light. Since CMB cannot provide a good constraint on DDME, we use DESI and Pantheon+ to break the parameter degeneracy. We find that the universe exhibits a similar behavior of matter evolution to that in CPL DDE-only model \cite{Wang:2025owe}. However, it only experiences a double acceleration during its evolution and ultimately reaches the Super Rip at beyond $1\,\sigma$ level. It is noteworthy that, with increasing the precision of parameters, CDP also suffers from tensions among the observations of CMB, DESI and Pantheon+.

{\it Special example.} As a special case of CPL DDM, $\Lambda$CDM with a varying constant DM EoS $\omega_{dm0}$ (hereafter $\Lambda\omega$DM) gives interesting fits to current observations (see SM). Note that constraints on $\Lambda\omega$DM have been implemented in \cite{Muller:2004yb,Calabrese:2009zza,Xu:2013mqe,Thomas:2016iav,Kopp:2018zxp} using early-time cosmological data. Here we find that C, D and P give $\omega_{dm0}=-0.0015\pm 0.0018, \, -0.030\pm 0.044$ and $-0.16^{+0.17}_{-0.15}$ indicating a $\sim1\,\sigma$ hint of negative pressure DM. DP provides $\omega_{dm0}=-0.038^{+0.025}_{-0.032}$ suggesting a $1.34\,\sigma$ preference of $\omega_{dm0}<0$ from low-$z$ data, while CP gives $\omega_{dm0}=-0.00174\pm 0.00089$ implying a $1.34\,\sigma$ deviation from $\Lambda$CDM. Furthermore, CD and CDP gives $\omega_{dm0}=0.00078\pm 0.00037$ and $0.00067\pm 0.00037$ indicating a $1.96\,\sigma$ and $1.81\,\sigma$ evidence of $\omega_{dm0}>0$, respectively. This is similar as expected in the DDME model because a larger $H_0$ induced by the combination of DESI BAO measurements and the derived $r_d$ from CMB. One can easily find that considering an evolving DM leads to the fact the constraint on today's DM EoS $\omega_{dm0}$ in DDM is clearly weakened by $\sim2$ orders compared to that in $\Lambda\omega$DM.

{\it Concluding remarks.} In the DDM, SDDM and DDME models, we find: (i) the robust $\sim2\,\sigma$ evidences of evolving DM using using individual datasets including CMB, DESI and Pantheon+; (ii) beyond $2\,\sigma$ signals of the coexistence of DDM and DDE using the data combinations of CD, CP and CDP. We cannot detect cosmic acceleration with any independent dataset in the DDM, SDDM and DDME scenarios as well as the CPL DDE model \cite{Wang:2025owe}. These models allow the possibility that current universe is undergoing cosmic deceleration or moving at a constant speed. More interestingly, due to their weak constraining power, individual datasets CMB, DESI and Pantheon+ in three models cannot demonstrate the existence of DE in both the present and the past, unlike in the DDE model where they can at least confirm the existence of DE at a beyond $5\,\sigma$ level \cite{Wang:2025owe}. Different from the Big Rip in the phantom energy model \cite{Caldwell:2003vq}, three evolving DM models prefer the Super Rip, which predicts that the universe is ultimately in the phase of super acceleration induced by DM with an extremely negative pressure.     

In theory, the DM temperature can be approximately as $T_{dm}(a)\approx\omega_{dm}(a)m_{dm}$ \cite{Armendariz-Picon:2013jej}, where $m_{dm}$ is the DM particle mass. Given the physically positive $m_{dm}$, our results prefer that the universe ends up with an infinitely negative DM temperature. This possibly results in a colder universe than that predicted by the Big Freeze \cite{Thomson1852}.
When $\omega_{dm}(a)<-1$, the absolute value of sound speed of DM would be greater than the speed of light $c$. This contradicts the principle of special relativity \cite{Einstein:1905vqn}, which states that no information or energy can travel faster than the speed of light.

{\it Note added}. During the preparation of this study, a paper \cite{Kumar:2025etf} appears on arXiv, which claims an evidence of non-zero DM EoS in the $\Lambda\omega$DM using the combination of DESI and other datasets. It is a special example in our analysis. However, we stress that this model cannot explain the true nature of DM on cosmic scales, starting from the viewpoint of DM EoS. Actually, the universe exhibits a stronger preference for DDM in light of current observations.

{\it Acknowledgements.} DW is supported by the CDEIGENT fellowship of Consejo Superior de Investigaciones Científicas (CSIC).

\clearpage

\appendix

\onecolumngrid
\section{\large Supplementary Material for ``Evidence for Dynamical Dark Matter''}
\twocolumngrid

\onecolumngrid

\setcounter{equation}{0}
\setcounter{figure}{0}
\setcounter{table}{0}

\section*{A. Perturbation formula}
Considering DM as a perfect fluid on cosmic scales, in the synchronous gauge \cite{Mukhanov:1992,Ma:1995,Malik:2009}, its perturbation evolution of overdensity and velocity divergence can be expressed as
\begin{align}
\dot{\delta}_{dm}&=-(1+\tilde{\omega}_{dm})(\theta_{dm}+\frac{\dot{h}}{2})+\frac{\dot{\tilde{\omega}}_{dm}}{1+\tilde{\omega}_{dm}}\delta_{dm}-3\mathcal{H}(c^2_{s,eff}-c^2_{s,ad})\left[\delta_{dm}+3\mathcal{H}(1+\tilde{\omega}_{dm})\frac{\theta_{dm}}{k^2}\right],  \label{eqs1}  \\                                              
\dot{\theta}_{dm}&=-\mathcal{H}(1-3c^2_{s,ad})\theta_{dm}+\frac{k^2c^2_{s,eff}}{1+\tilde{\omega}_{dm}}\delta_{dm}-k^2\sigma_{dm}, \label{eqs2}   \end{align}
where $\tilde{\omega}_{dm}\equiv\omega_{dm}(a)$, $\mathcal{H}\equiv aH$ is the conformal Hubble parameter, $c^2_{s,ad}\equiv \tilde{\omega}_{dm}-\dot{\tilde{\omega}}_{dm}/[3\mathcal{H}({1+\tilde{\omega}_{dm}})]$ is the adiabatic sound speed of DM, $c^2_{s,eff}$ is the effective sound speed of DM, a free parameter characterizing the small scale behaviors of DM, and $\sigma_{dm}$ denotes the shear perturbation of DM depicting the anisotropic stress of DM (see \cite{Hu:1998kj} for details). In our analysis, we adopt the standard CPL DDE perturbation used by the Planck collaboration \cite{Planck:2018vyg1}.

\begin{figure}[h!]
	\centering
	\includegraphics[scale=0.5]{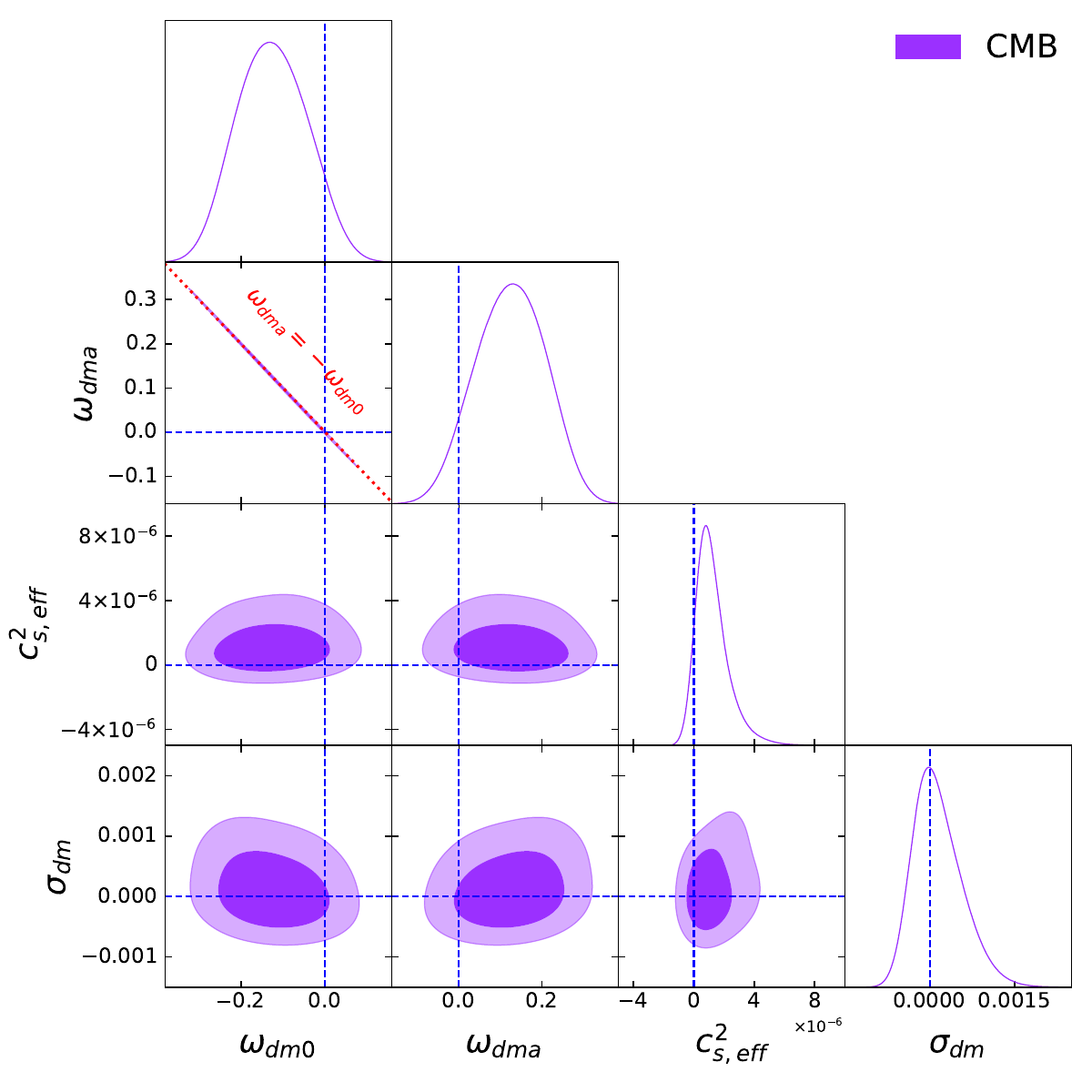}
	\caption{One-dimensional and two-dimensional posterior distributions of four free parameters from the CMB dataset in the extended DDM model. The blue dashed lines are $\omega_{dm0}=\omega_{dma}=c^2_{s,eff}=\sigma_{dm}=0$, while the red dotted line denotes the linear relation $\omega_{dma}=-\omega_{dm0}$.}\label{fs1}
\end{figure}

\begin{figure}
	\centering
	\includegraphics[scale=0.5]{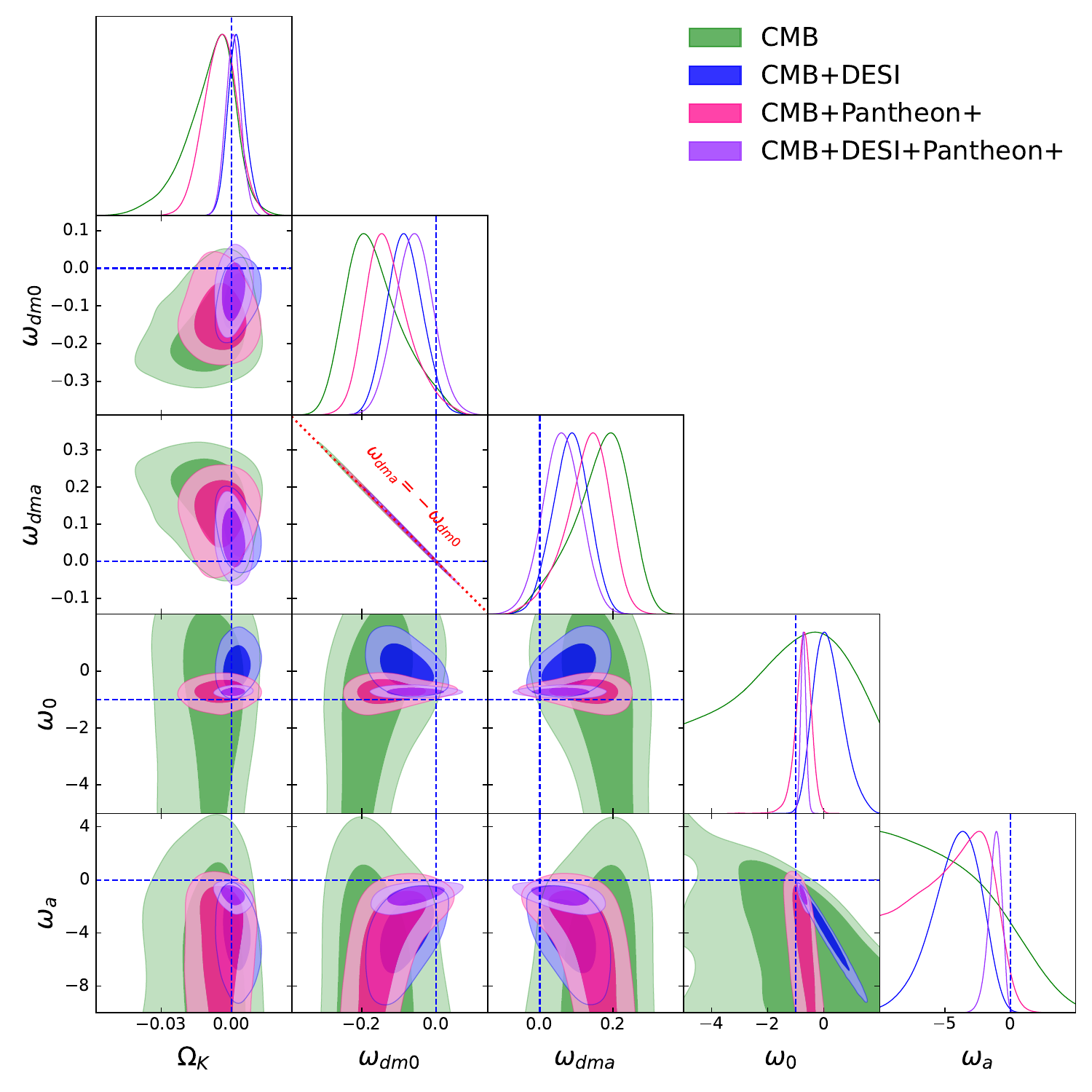}
	\caption{One-dimensional and two-dimensional posterior distributions of the main parameters from various datasets in the DDME model with a free cosmic curvature. The blue dashed lines are $\Omega_K=\omega_{dm0}=\omega_{dma}=\omega_a=0$ and $\omega_0=-1$, while the red dotted line denotes the linear relation $\omega_{dma}=-\omega_{dm0}$.}\label{fs2}
\end{figure}

\begin{figure}[h!]
	\centering
	\includegraphics[scale=0.55]{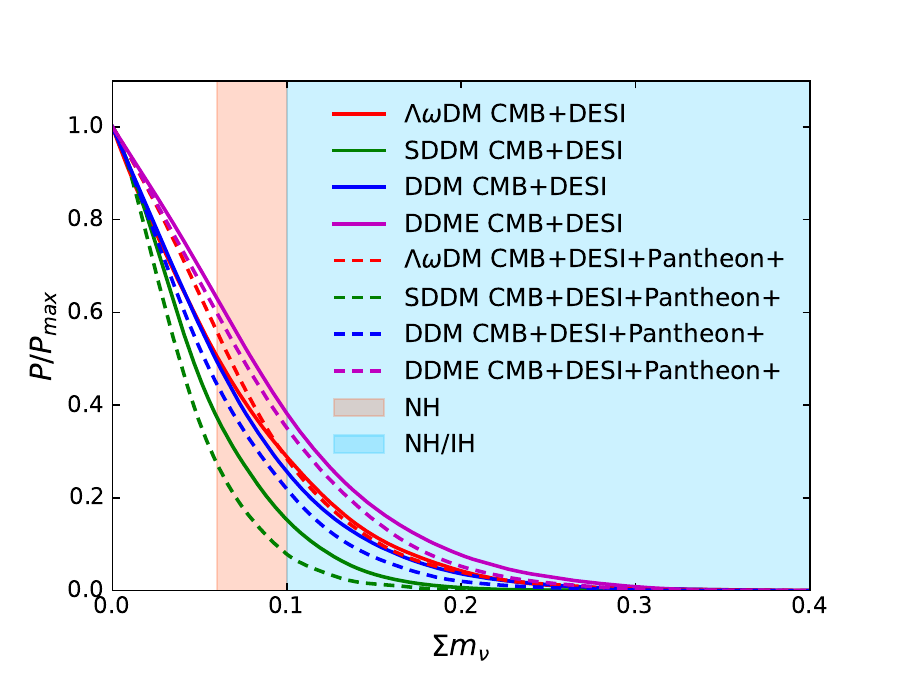}
	\caption{One-dimensional posterior distributions of the parameter $\Sigma m_\nu$ from different datasets in the $\Lambda\omega$DM, SDDM, DDM and DDME models.}\label{fs3}
\end{figure}

\section*{B. The four-parameter space of DDM}
In the main text, we only show the constraints on the parameters of DM EoS in the DDM, SDDM and DDME models, because the other two parameters $c^2_{s,eff}$ and $\sigma_{dm}$ do not affect the constraints on the DM EoS. Here we explore the  four-parameter space of evolving DM using the CMB observations. We obtain the $1\,\sigma$ constraints $\omega_{dm0}=-0.126\pm 0.082$, $\omega_{dma}=0.125\pm 0.083$, $c^2_{s,eff}=\left(\,1.18^{+0.70}_{-1.20}\,\right)\times 10^{-6}$ and $\sigma_{dm}=0.00013^{+0.00034}_{-0.00049}$ (see Fig.\ref{fs1}). This means that the DDM evidence is reduced from $1.92\,\sigma$ to $1.73\,\sigma$ when considering the effective sound speed and anisotropic stress of DM. We think that this reduction can be mainly ascribed to the fact that the addition of $c^2_{s,eff}$ and $\sigma_{dm}$ leads to the enlargement of the ($\omega_{dm0},\,\omega_{dma}$) parameter space. Moreover, we do not find any signal of non-zero effective sound speed and anisotropic stress of DM. The very strong linear relation $\omega_{dma}=-\omega_{dm0}$ is well confirmed in the extended DDM parameter space. 

\begin{table*}[h!]
	\renewcommand\arraystretch{1.5}
	\caption{Mean values and $1\,\sigma$ (68\%) errors of free parameters from various datasets in the DDME model. We quote $2\,\sigma$ (95\%) upper limits for parameters with weak constraints, while using the symbols ``$\bigstar$'' to denote unconstrained parameters by data.}
	\setlength{\tabcolsep}{5mm}{
		\begin{tabular} { c |c| c |c| c|c}
			\hline
			\hline
			
			\multicolumn{2}{c|}{Data}                & C    & CD    & CP & CDP                  \\
			\hline
			\multirow{5}{1cm}{DDME}  &     $\omega_{dm0}$  & $-0.164^{+0.055}_{-0.092}  $ & $-0.084\pm 0.047           $ & $-0.129^{+0.045}_{-0.069}  $ & $-0.059\pm 0.051           $   \\
			\cline{2-6}
			&     $\omega_{dma}$  & $0.164^{+0.093}_{-0.057}   $ & $0.085\pm 0.049            $ & $0.129^{+0.070}_{-0.047}   $ & $0.061\pm 0.052            $\\
			\cline{2-6}                        
			&     $\omega_{0}$   & $\bigstar        $ & $0.17^{+0.43}_{-0.60}      $ & $-0.74^{+0.30}_{-0.23}     $ & $-0.725\pm 0.095           $\\
			\cline{2-6}	                        		                        
			&     $\omega_{a}$    & $< 1.85                  $ & $-4.2^{+2.2}_{-1.5}        $ & $< -0.660        $ & $-1.19^{+0.58}_{-0.40}     $\\
			\cline{2-6}
			&     $\Omega_{K}$ & $-0.0094^{+0.0120}_{-0.0073}$ & $0.0022^{+0.0034}_{-0.0042}$ & $-0.0049\pm 0.0070         $ & $0.0006\pm 0.0033          $ \\
			
			\hline
			\hline
	\end{tabular}}
	\label{ts1}
\end{table*}

\begin{table*}[h!]
	\renewcommand\arraystretch{1.6}
	\caption{Mean values and $1\,\sigma$ (68\%) errors of the sum of neutrino masses $\Sigma m_\nu$ from different datasets in the $\Lambda\omega$DM, SDDM, DDM and DDME models. We quote $2\,\sigma$ (95\%) upper limits for $\Sigma m_\nu$. }
	\setlength{\tabcolsep}{10mm}{
		\begin{tabular} { c |l| c }
			\hline
			\hline
			
			\multicolumn{2}{c|}{Data}                & $\Sigma m_\nu$ [eV]                    \\
			\hline
			\multirow{2}{1cm}{$\Lambda\omega$DM}  & CD  &$< 0.168 $                                                \\
			\cline{2-3}
			&    CDS  & $< 0.162   $             \\
			\hline
			\multirow{2}{1cm}{SDDM}  & CD  &$<0.120  $                                                \\
			\cline{2-3}
			&    CDS  & $< 0.0994    $             \\
			\hline
			\multirow{2}{1cm}{DDM}  & CD  &$< 0.164$                                                \\
			\cline{2-3}
			&    CDS  & $< 0.145   $             \\
			\hline
			\multirow{2}{1cm}{DDME}  & CD  &$< 0.187$                                                \\
			\cline{2-3}
			&    CDS  & $< 0.170   $             \\
			
			\hline
			\hline
	\end{tabular}}
	\label{ts2}
\end{table*}

\section*{C. The effect of curvature on the dynamical dark sector}
The cosmic curvature $\Omega_K$ plays an important role in depicting the background evolution and structure formation of the universe. Here we study its impact on the coexistence of DDM and DDE in the DDME scenario. Overall, we find that $\Omega_K$ hardly affect our conclusions. Interestingly, CMB gives $\omega_{dma}=0.164^{+0.093}_{-0.057}$, which is a $2.53\,\sigma$ evidence for the redshift evolution of DM (see Fig.\ref{fs2} and Tab.\ref{ts1}). The addition of $\Omega_K$ increases the DDM evidence from $1.82\,\sigma$ to $2.53\,\sigma$. CD, CP and CDP provides a $1.93\,\sigma$, $2.40\,\sigma$ and $1.31\,\sigma$ evidence of DDM, respectively. Furthermore, we find that C, CD, CP and CDP exhibit a $\sim2\,\sigma$, $2.37\,\sigma$, beyond $2\,\sigma$, and $2.40\,\sigma$ signal of DDE, respectively. It is worth noting that in the planes of $\omega_{dm0}-\omega_0$, $\omega_{dm0}-\omega_a$, $\omega_{dma}-\omega_0$ and $\omega_{dma}-\omega_a$, C, CD, CP and CDP still support a beyond $2\,\sigma$ evidence of the coexistence of DDM and DDE in the dark sector of the universe. Additionally, the $\sim2\,\sigma$ preference of a closed universe by Planck CMB data could be alleviated when simultaneously considering the DDM and DDE.

\begin{figure}[h!]
	\centering
	\includegraphics[scale=0.5]{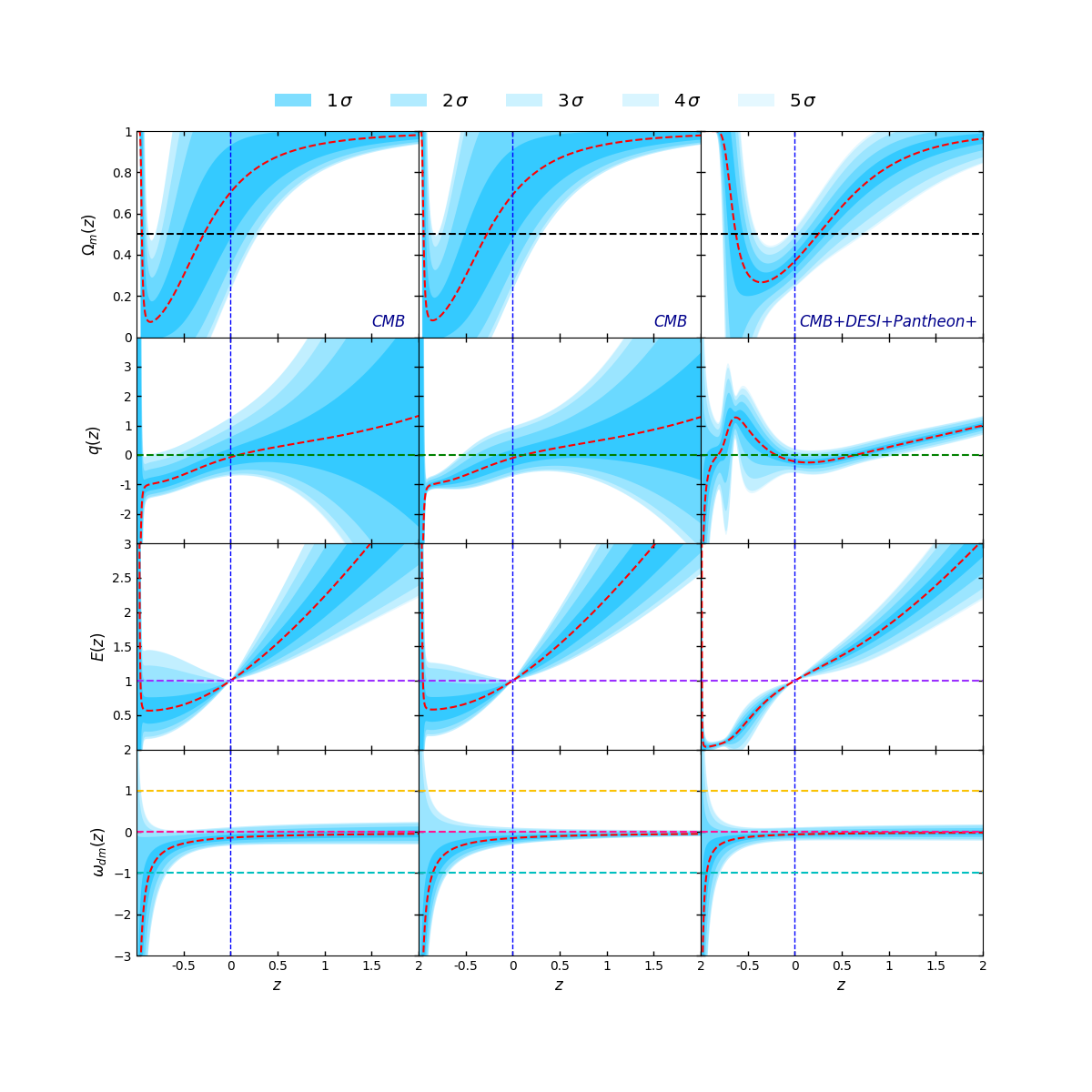}
	\caption{The $1\,\sigma-5\,\sigma$ regions of background quantities $\Omega_m(z)$, $q(z)$, $E(z)$ and $\omega_{dm}(z)$ from different datasets in the DDM (left), SDDM (middle) and DDME (right) models. The dashed lines denote the mean values of each quantity (red), $z=0$ (blue), $\Omega_m=0.5$ (black), $q=0$ (green), $E=1$ (purple), $\omega_{dm}=0$ (pink), $\omega_{dm}=1$ (orange) and $\omega_{dm}=-1$ (cyan), respectively.}\label{fs4}
\end{figure}

\begin{figure}[h!]
	\centering
	\includegraphics[scale=0.5]{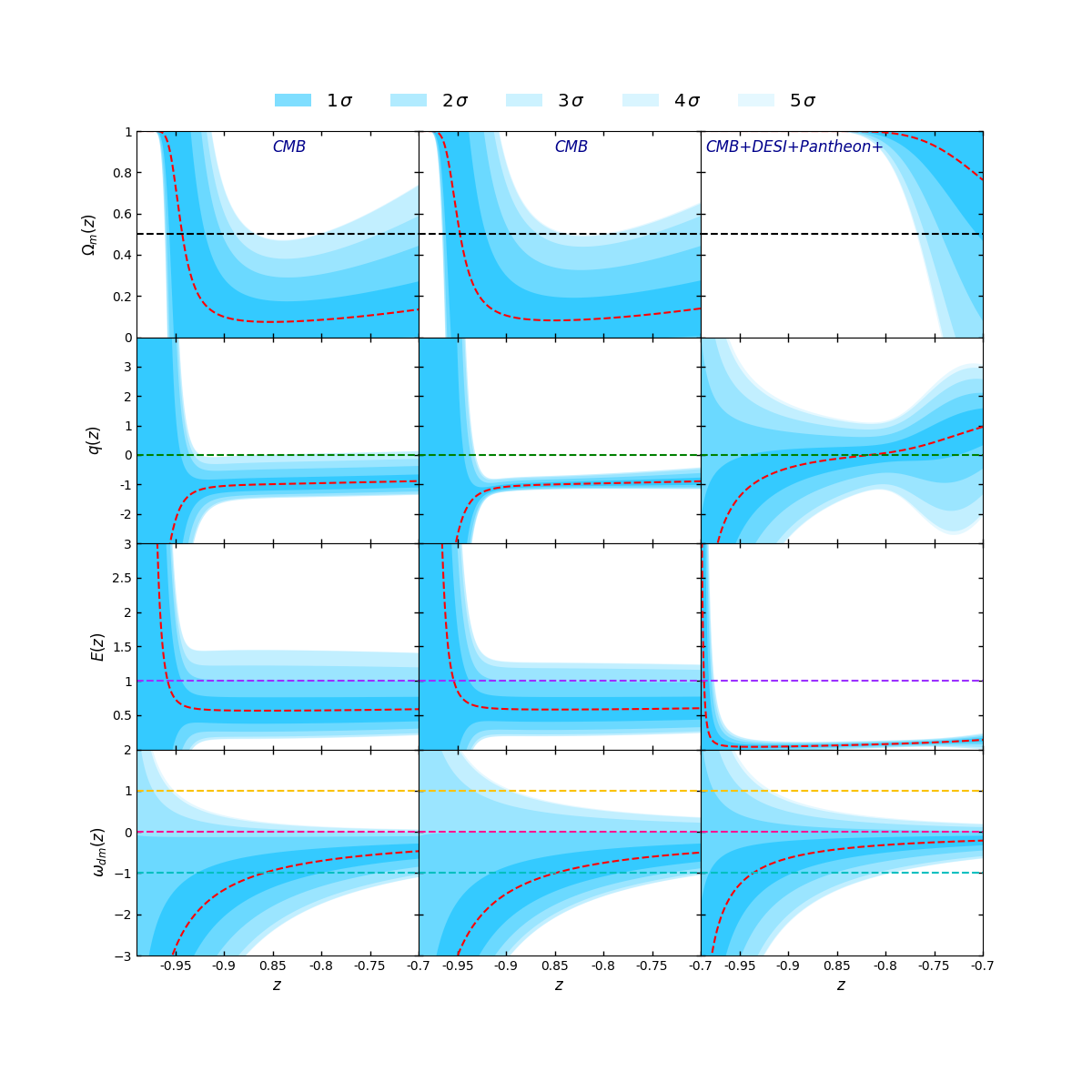}
	\caption{Same as Fig.\ref{fs3} but in the zoomed redshift range $z\in[-1,-0.7]$.}\label{fs5}
\end{figure}

\begin{table*}[h!]
	\renewcommand\arraystretch{1.6}
	\begin{center}
		\caption{Mean values and $1\,\sigma$ (68\%) errors of free parameters from different datasets in the $\Lambda\omega$DM model. We quote $2\,\sigma$ (95\%) upper limits for parameters with weak constraints. }
		\setlength{\tabcolsep}{10mm}{
			\begin{tabular}{l |c }
				\hline
				\hline
				Parameter & $\omega_{dm0}$    \\
				\hline 
				CMB &  $-0.0015\pm 0.0018 $                     \\
				CMB+DESI &  $0.00078\pm 0.00037$        \\
				CMB+Pantheon+ &  $-0.00174\pm 0.00089$    \\
				CMB+DESI+Pantheon+ &  $0.00067\pm 0.00037$      \\
				\hline
				DESI &  $-0.030\pm 0.044$               \\
				Pantheon+ &  $-0.16^{+0.17}_{-0.15}$     \\
				Union3 &  $-0.27^{+0.14}_{-0.19}$      \\
				DESY5 &  $-0.23\pm 0.17$      \\
				DESI+Pantheon+ &  $-0.038^{+0.025}_{-0.032}$   \\
				DESI+Union3 &  $-0.058\pm 0.029$     \\
				DESI+DESY5 &  $-0.063\pm 0.024 $     \\
				
				\hline
				\hline
		\end{tabular}}
		\label{ts3}
	\end{center}
\end{table*} 

\section*{D. Relieving neutrino mass tension with DDM}
Employing the degeneracy between DE and neutrino masses, the DESI collaboration demonstrate that CPL DDE can help relax the $2\,\sigma$ upper limit to a safe region \cite{DESI:2025ejh}. Here we verify that evolving DM can also well resolve the neutrino mass tension between cosmological observations and terrestrial experiments (see Fig.\ref{fs3} and Tab.\ref{ts2}). Using the combination of CMB and DESI data, we give, respectively, the $2\,\sigma$ upper bound $\Sigma m_\nu<0.168$, 0.120, 0.164 and 0.187 eV in the $\Lambda\omega$DM, SDDM, DDM and DDME models. If adding the Pantheon+ SN data, the bounds are slightly tightened to $\Sigma m_\nu<0.162$, 0.0994, 0.145 and 0.170 eV. This implies that there exists a strong degeneracy between the DM EoS and neutrino mass, similar to the case of DE.

\section*{E. Destiny+}
In \cite{Wang:2025owe}, based on the latest cosmological observations, we have proposed the new fate of the universe called the ``Big Stall'' in the CPL DDE model. In this work, when considering the coexistence of DDM and DDE, we propose a new density of the universe, i.e., the ``Super Rip'' in the DDME model. They both support that the universe is ultimately dominated by DM rather than DE. In Figs.\ref{fs4} and \ref{fs5}, we show the $1\,\sigma-5\,\sigma$ regions of background quantities $\Omega_m(z)$, $q(z)$, $E(z)$ and $\omega_{dm}(z)$ from different datasets in the DDM, SDDM and DDME scenarios. The details of their properties and the Super Rip have been specified in the main text.

\begin{figure}[h!]
	\centering
	\includegraphics[scale=0.55]{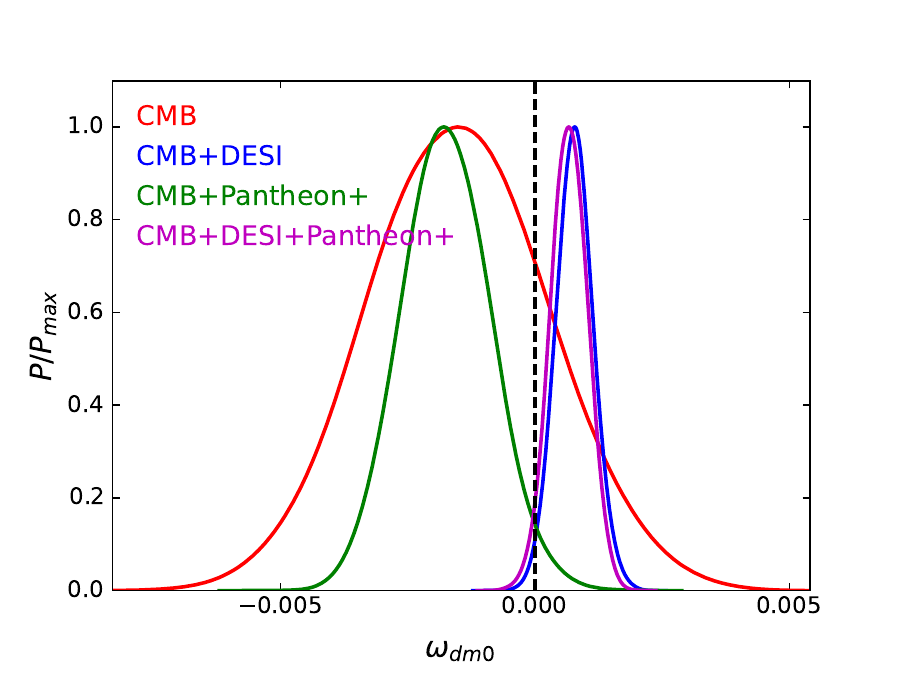}
	\includegraphics[scale=0.55]{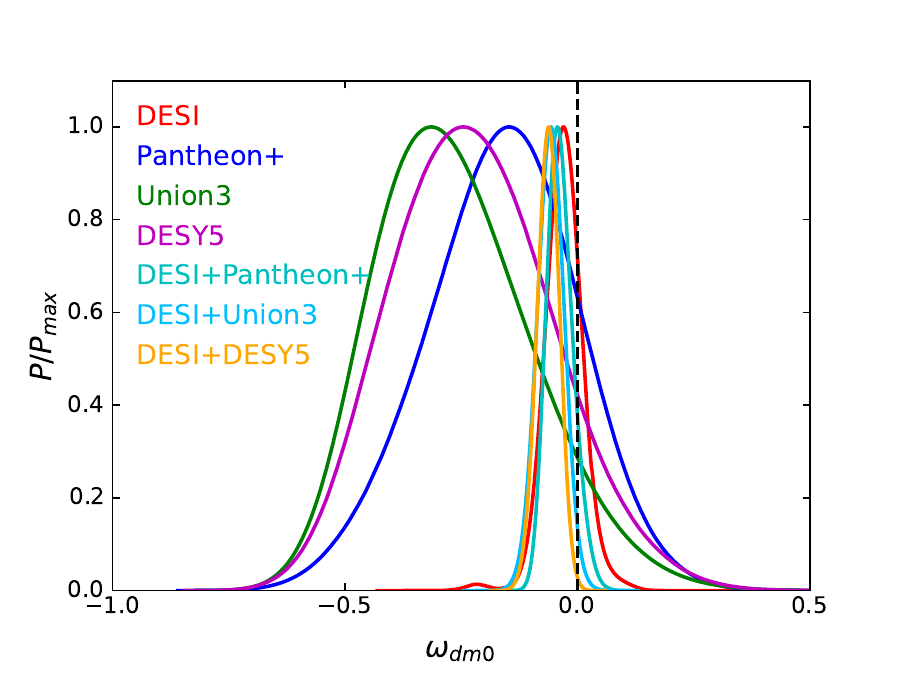}
	\caption{One-dimensional posterior distributions of the parameter $\omega_{dm0}$ from CMB-related (left) and late-time (right) datasets in the $\Lambda\omega$DM model. The black dashed line denotes the $\Lambda$CDM model.}\label{fs6}
\end{figure}

\section*{F. Special example: constraints on $\omega_{dm0}$}
In the main text, we just show results from the Pantheon+ SN dataset. As a comparison, here we also consider two extra SN samples: (i) Union3 with 22 spline-interpolated data points derived by 2087 SN from 24 different surveys in $z\in[0.05, 2.26]$ \cite{Rubin:2023ovl}; (ii) DESY5 consisting of 1735 effective data points in $z\in[0.025, 1.130]$ \cite{DES:2024jxu}. In Fig.\ref{fs6} and Tab.\ref{ts3}, as mentioned in the main text, it is easy to see that both high-$z$ CMB and low-$z$ BAO and SN datasets independently prefer a negative $\omega_{dm0}$. However, it is important to note that this model does not fully account for the true nature of DM on cosmic scales, particularly when considering the dynamical properties of DM. Actually, the universe exhibits a stronger preference for DDM in light of the latest observations.

\begin{figure}[h!]
	\centering
	\includegraphics[scale=0.55]{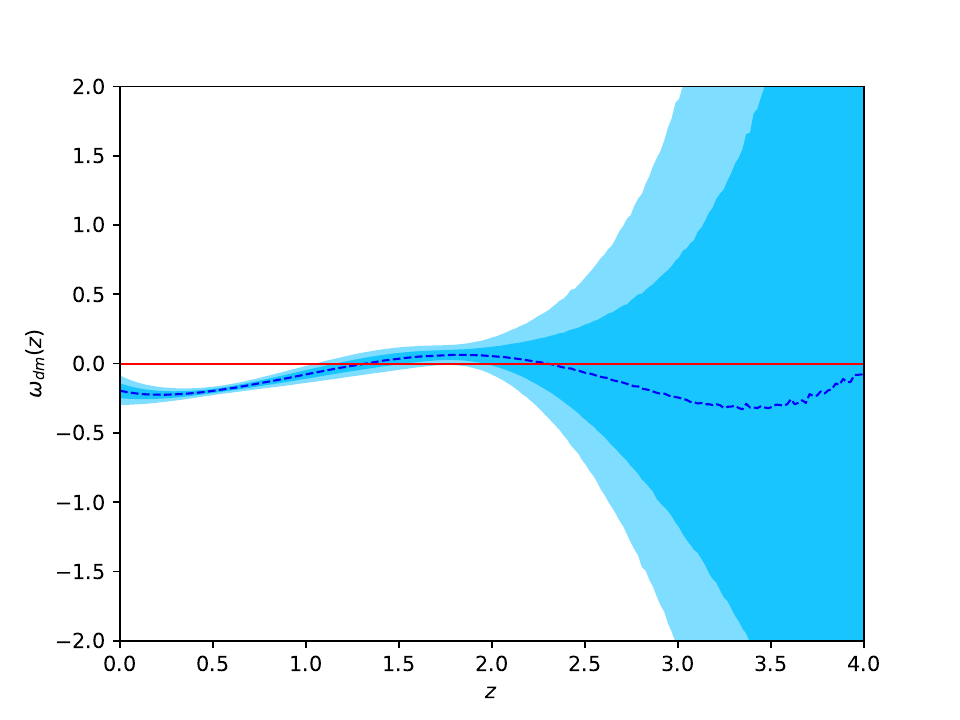}
	\caption{The GP reconstructions of the DM EoS by using the combination of CMB, DESI, Pantheon+ and CC datasets, respectively. The shaded regions denote the $1\,\sigma$ (dark) and $2\,\sigma$ (light) uncertainties of reconstructed DM EoS. The solid (red) and dashed (blue) lines represent the $\Lambda$CDM model and the underlying DM EoS from data, respectively.}\label{fs7}
\end{figure}

\section*{G. Reconstructing $\omega_{dm}(z)$ with Guassian process}
It is clear that our results are model-dependent. Here we show a new possibility of exploring the redshift evolution of the DM EoS in a model-independent way. Specifically, we use the Gaussian process (GP) (see \cite{Seikel:2012uu} for more details) to implement a model-independent reconstruction of the DM EoS over time. We take the Pantheon+, DESI BAO, cosmic chronometers (CC) \cite{Moresco:2016mzx} and the shift parameter $\mathcal{R}(z)$ derived from the CMB data to implement the GP reconstruction.

In a homogeneous and isotropic universe, the luminosity distance $D_L(z)$ of SN is written as
\begin{equation}
D_L(z)=\frac{c\,(1+z)}{H_0\sqrt{|\Omega_{k}|}}\mathrm{sinn}\left(\sqrt{|\Omega_{k}|}\int^{z}_{0}\frac{dz'}{E(z')}\right), \label{eqs3}
\end{equation}
where the dimensionless Hubble parameter $E(z)\equiv H(z)/H_0$, today's cosmic curvature $\Omega_{K}=-\mathrm{K}c^2/(H_0^2)$, and for $\mathrm{sinn}(x)= \mathrm{sin}(x), x, \mathrm{sinh}(x)$, $\mathrm{K}=1, \, 0, \, -1$ , which corresponds to a closed, flat and open universe, respectively. Employing the normalized comoving distance $D(z)=(H_0/c)(1+z)^{-1}D_L(z)$, the DM EoS reads as
\begin{equation}
\omega_{dm}(z)=\frac{2D''(1+z)-3D'\left[(1-\Omega_m)D'^2-1\right]}{3D'\left[\Omega_b(1+z)^3D'^2+(1-\Omega_m)D'^2-1\right]}, \label{eqs4}
\end{equation}
where the prime denotes the derivative with respect to the redshift $z$. The background quantities $D'$ and $D''$ in Eq.(\ref{eqs4}) can be easily obtained using the GP reconstructions. In our analysis, for simplicity, we consider a flat universe, namely $\Omega_{k}=0$ and assume $H_0=73.04\pm1.04$ km s$^{-1}$ Mpc$^{-1}$ \cite{Riess:2021jrx}. We also take $\Omega_{m}=0.3153\pm0.0073$ and $\Omega_{b}=0.04932\pm0.00052$ \cite{Planck:2018vyg1}. For DESI BAO data, assuming $r_d=147.09\pm0.26$ Mpc, we transform $D_M(z)/r_d$ and $D_H(z)/r_d$ to $D$ and $D'$, respectively. For CC, we directly transform $H(z)$ to $D'$, while transforming $D_L(z)$ from SN and $\mathcal{R}(z_c)=1.7488\pm0.0074$ \cite{Planck:2015bue} ($z_c=1089.0$ is the redshift of recombination) from CMB to $D$. Same as \cite{Seikel:2012uu}, we use the so-called Mat\'{e}rn ($\nu=9/2$) covariance function in GP reconstructions. Based on Eq.(\ref{eqs4}) and different kinds of datasets, we use a modified version of the package \texttt{GaPP} to carry our the GP reconstructions.

\begin{figure}[h!]
	\centering
	\includegraphics[scale=0.55]{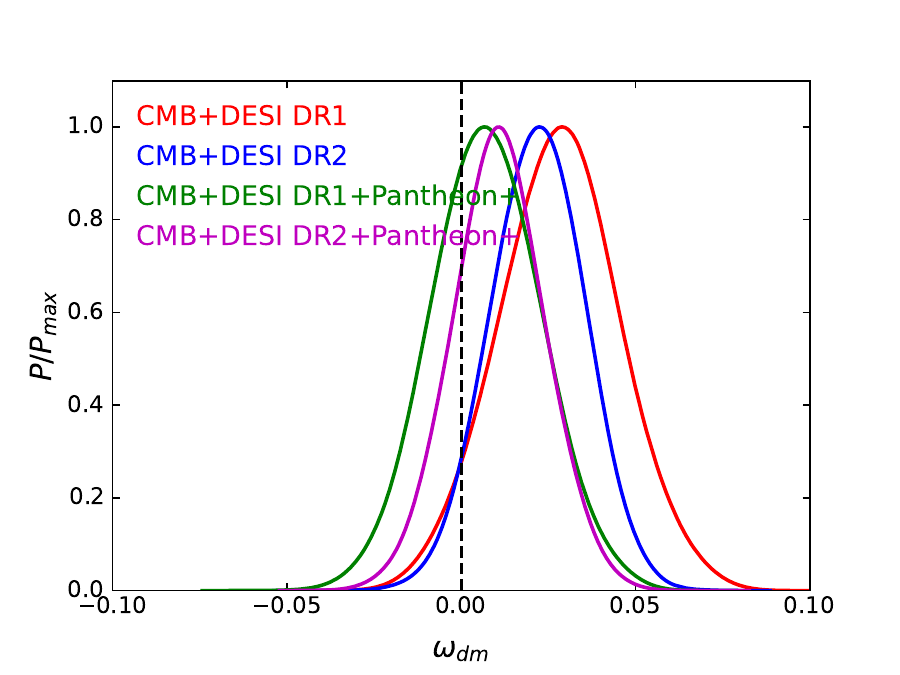}
	\caption{One-dimensional posterior distributions of the parameter $\omega_{dm}$ from different datasets in the SDDM model. The black dashed line denotes the $\Lambda$CDM model.}\label{fs8}
\end{figure}

\begin{figure}[h!]
	\centering
	\includegraphics[scale=0.55]{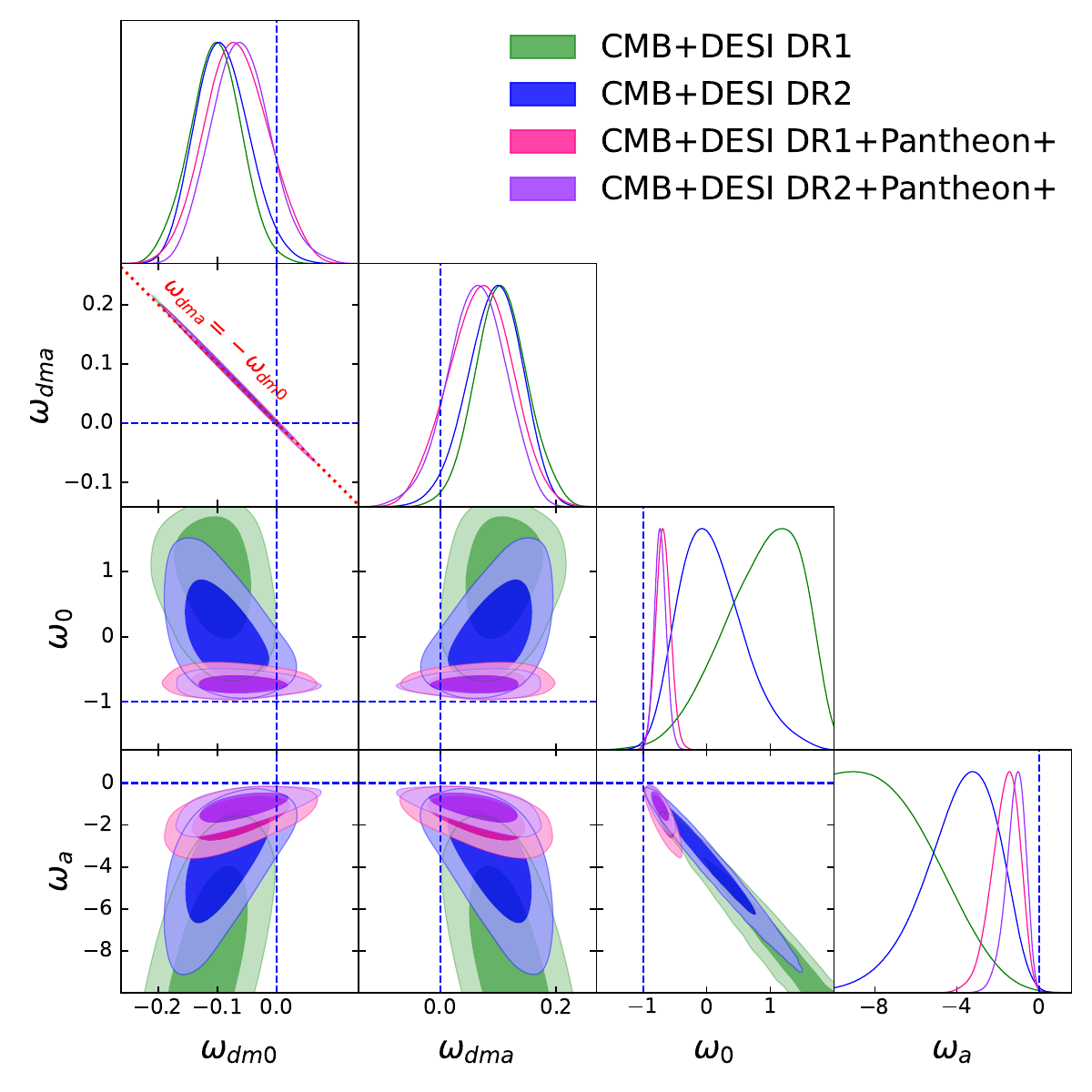}
	\caption{One-dimensional and two-dimensional posterior distributions of the main parameters from various datasets in the DDME model. The blue dashed lines are $\omega_{dm0}=\omega_{dma}=\omega_a=0$ and $\omega_0=-1$, while the red dotted line denotes the linear relation $\omega_{dma}=-\omega_{dm0}$.}\label{fs9}
\end{figure}

In Fig.\ref{fs7}, it is easy to see that: (i) When $z>2$, the constraining power decreases rapidly due to the limitation of redshift distribution of data; (ii) When $z\approx1.8$, there is a $2\,\sigma$ evidence of $\omega_{dm}(z)>0$; (iii) Starting from $z\lessapprox1$, observations exhibit a beyond $2\,\sigma$ signal of $\omega_{dm}(z)<0$ until today. This implies that there exist a $2\,\sigma$ evidence for the zero-crossing DM EoS, which is warm around $z=1.8$, then become cold and finally produces a negative pressure DM today. Interestingly, in light of the latest observations, both GP reconstructions and DDM, SDDM and DDME models support that today's DM EoS has a negative pressure ($\omega_{dm}(z=0)<0$) at the $2\,\sigma$ confidence level.

\section*{H. Comparison of DESI DR1 and DR2}
In Figs.\ref{fs8} and \ref{fs9}, we demonstrate the robustness of our conclusions and show the improvements of the constraining power of DESI DR2 BAO measurements relative to DESI DR1 data in the SDDM and DDME models. One can easily find that the constraints from DESI DR1 and DR2 data on evolution of DDM is robust and that the high-precision DESI DR2 BAO data significantly compresses the parameter space of DDE.

\end{document}